\documentclass[aps,prl,
reprint, superscriptaddress
]{revtex4-1}
\usepackage[margin=0.6in]{geometry}
\usepackage{amsthm,amsmath,physics,mathtools,amssymb}

\usepackage{charter}

\usepackage{CJKutf8}

\usepackage{enumitem}

\theoremstyle{definition}

\usepackage{graphicx}
\graphicspath{ {images/} }

\usepackage{tensor}

\DeclarePairedDelimiterX{\inp}[2]{\langle}{\rangle}{#1, #2}

\DeclareMathOperator{\sinc}{sinc}

\usepackage{hyperref}
\hypersetup{
    colorlinks=true,
    linkcolor=blue,
    filecolor=magenta,      
    urlcolor=cyan,
}

\usepackage{cleveref}

\usepackage{xparse}

\NewDocumentCommand\LH{mo}{%
  \IfNoValueTF{#2}
   {\mathcal{L}(\mathcal{H}^{#1})}
   {\mathcal{L}(\mathcal{H}^{#1},\mathcal{H}^{#2})}%
}

\newcommand\id{\leavevmode\hbox{\small1\kern-3.3pt\normalsize1}}

\allowdisplaybreaks

\begin{document}

\begin{CJK*}{UTF8}{gbsn}
\title{Ultraviolet regularity from quantum gravitational indefinite causal structure}
\author{Ding Jia (贾丁)}
\email{ding.jia@uwaterloo.ca}
\affiliation{Perimeter Institute for Theoretical Physics, Waterloo, Ontario, N2L 2Y5, Canada}
\affiliation{Department of Physics and Astronomy, University of Waterloo, Waterloo, Ontario, N2L 3G1, Canada}
\begin{abstract}
Indefinite causal structure is generically present in theories of quantum gravity admitting a path integral formulation. We show that summing over causal structures eliminates ultraviolet divergences of matter QFT and resolves spacetime singularities using the non-perturbative World Quantum Gravity approach. Independent information-theoretic and model-independent considerations suggest that the mechanism of ultraviolet regularization by indefinite causal structure also applies to other theories of quantum gravity.
\end{abstract}
\maketitle
\end{CJK*}

\section{Introduction}

\begin{figure}
    \centering
    \includegraphics[width=.35\textwidth]{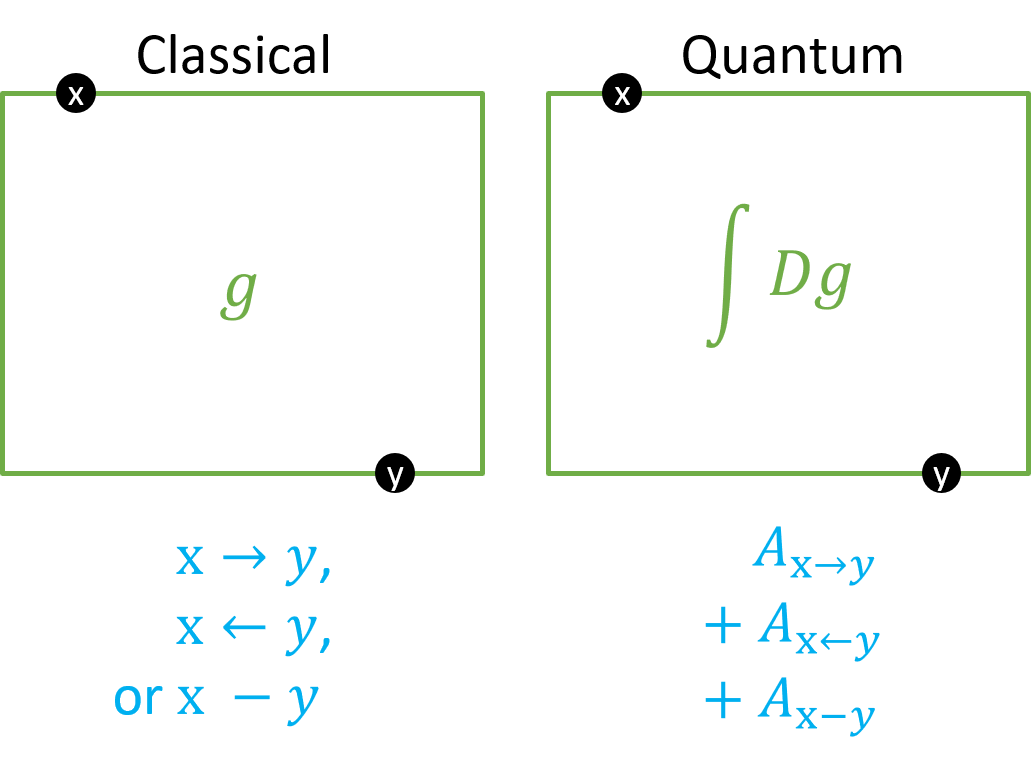}
    \caption{Indefinite causal structure in quantum gravitational path integrals}
    \label{fig:icspi}
\end{figure}

Indefinite causal structure \cite{HardyProbabilityGravity, hardy2007towards} is generically present in theories of quantum gravity that admit a path integral formulation (\Cref{fig:icspi}). Consider two events $x$ and $y$ identified in a physically meaningful way (e.g., by specifying the values certain scalar fields take at these events \cite{Westman2008EventsTheories, Westman2009CoordinatesRelativity, hardy2016operational}). On a classical spacetime configuration $g$, the events have one of the causal relations $x\rightarrow y$ ($x$ precedes $y$), $x\leftarrow y$ ($y$ precedes $x$), and $x-y$ ($x$ disconnected with $y$). On quantum spacetime, different configurations $g$ are summed over in a path integral to obtain the physical amplitudes $A=A_{x\rightarrow y}+A_{x\leftarrow y}+A_{x-y}$ that receive contributions from different causal structures. The causal structure for $x$ and $y$ is in this sense \textit{indefinite}.

In this letter, we study the effects of indefinite causal structure on ultraviolet (UV) physics, and show that it eliminates UV divergences of matter QFT and resolves spacetime singularities. 

Matter QFT UV divergences originate from short invariant distance artifacts \cite{Collins1984Renormalization}. Intuitively, it is reasonable that the quantum superposition over causal structures in a quantum gravitational path integral induces a smearing of the lightcone to cured the divergences. Indeed, similar ideas have been pursued in perturbative or partial considerations of quantum gravity \cite{deser1957, Ford1995GravitonsFluctuations, Padmanabhan1998HypothesisPropagator, Ohanian1999}. Here we extend the study to non-perturbative and full considerations of quantum gravity. 

Classical spacetime singularities originate from the gravitational focusing of non-spacelike curves  \cite{Hawking1973TheSpace-Time, wald2010general}. For example, in a Schwarzschild spacetime curves that had entered the event horizon cannot escape and will terminate at the singularity in finite time, \textit{unless the curves have spacelike parts}. It is again intuitively reasonable that quantum superposition over causal structures could avoid singularities, because then curves will not be strictly non-spacelike. 

To see how these intuitive expectations are realized explicitly in quantum gravity, we will need to work with a concrete theory. The recently introduced World Quantum Gravity \cite{JiaWorldFunction} is suitable for studying quantum gravitational indefinite causal structures. Ultraviolet regularity will be shown in a simple manner in this non-perturbative path integral approach. We then draw results from independent information-theoretic and model-independent studies to suggest that the mechanism of UV regularization by indefinite causal structure also applies to other theories of quantum gravity.

\section{World quantum gravity}

World quantum gravity (WQG) is a non-perturbative, Lorentzian, path integral approach to quantum gravity based on relational variables. The main formula (\ref{eq:mfp}) is arrived at by making the following choices \cite{JiaWorldFunction}:
\begin{itemize}
\item Adopt the path integral formulation of quantum theory.
\item Use the relational world function $\sigma(x,y)$ \cite{Synge1971Relativity:Theory} as a basic variable.
\item Use the ``worldline path integral'' representation of matter QFT  \cite{Feynman1950MathematicalInteraction, *Feynman1951AnElectrodynamics, CorradiniSpinningTheory}.
\item Use Parker's formula (\ref{eq:pmf2}) \cite{Parker1979PathSpace} to relate to the Einstein-Hilbert action.
\end{itemize}

The world function is one half the squared geodesic distance \cite{Synge1971Relativity:Theory}. For $ds^2 = g_{ab}dx^a dx^b$, $\sigma(x,y)=\frac{1}{2}\int_x^y ds^2$, where the integral is along the geodesic from $x$ to $y$ \footnote{It is assumed that $y$ is in the convex normal neighborhood of $x$ so the geodesic is unique.}. $\sigma$ contains all the information of the metric $g_{ab}$, as $g_{ab}(x)=-\lim_{y\rightarrow x}\frac{\partial}{\partial x^a}\frac{\partial}{\partial y^b}\sigma(x,y)$ \cite{Synge1971Relativity:Theory}.
The key idea of WQG is to \textit{express quantum gravity in terms of the relational world function $\sigma(x,y)$ instead of the metric field $g_{ab}(x)$}. Major motivations are: 1) $\sigma$ is \textbf{simple}. In particular, it is an invariant scalar; 2) $\sigma$ is \textbf{matter-friendly}. It provides a convenient way to incorporate matter QFT into the quantum gravity theory through the worldline path integral formalism; 3) $\sigma$ \textbf{indicates causal structure}. $\sigma<,=,>0$ manifestly correspond to time-, light-, and space-like separations. 

\begin{figure}
    \centering
    \includegraphics[width=.45\textwidth]{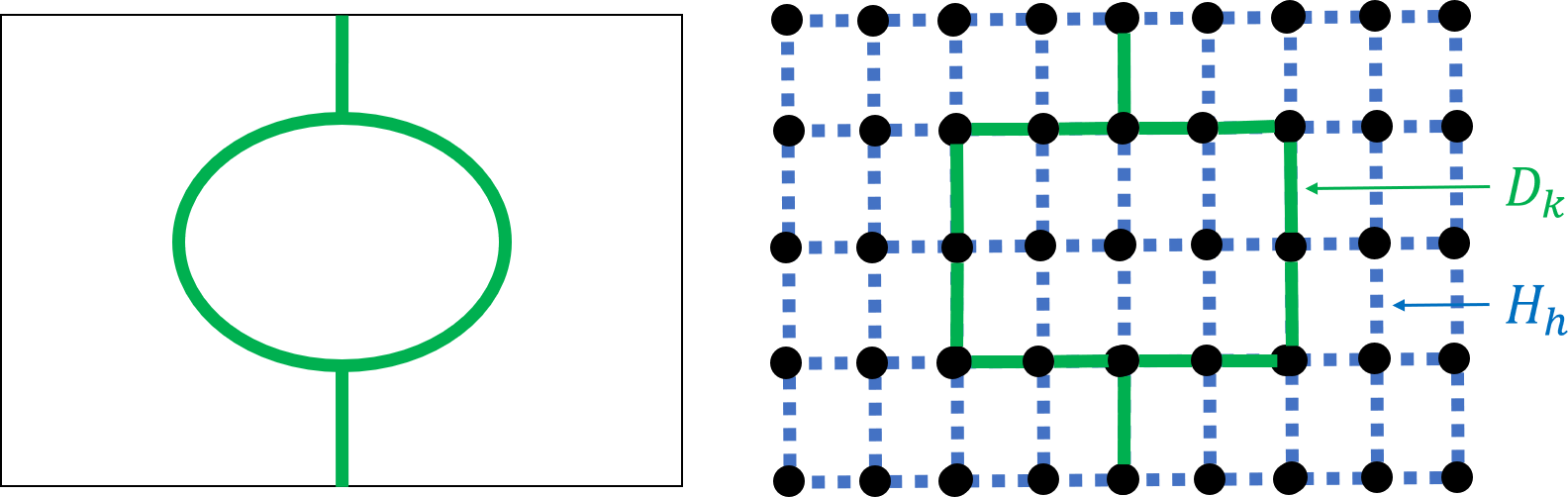}
    \caption{Traditional Feynman diagram vs. correlation diagram for WQG}
    \label{fig:wqgpic}
\end{figure}

The main result of \cite{JiaWorldFunction} is formula (\ref{eq:mfp}) for probability amplitudes of quantum processes with matter and gravity. 
Recall the ordinary Feynman rules for matter QFT \cite{Schwartz2013QuantumModel}
\begin{align}\label{eq:fr}
&\sum_\Gamma \prod_i \int dx_i \frac{V[\Gamma]}{N[\Gamma]}
\prod_{k\in \Gamma} G_{k},
\end{align}
where the sum is over Feynman diagrams $\Gamma$, the products are over vertices $i$ and edges $k$ of the diagrams, $V$ is the vertex factor, $N$ is the symmetry factor, and $G$ are the Feynman propagators.
The WQG formula (\ref{eq:mfp}) can be regarded as an improvement over (\ref{eq:fr}) to incorporate quantum gravity non-perturbatively. As illustrated in \Cref{fig:wqgpic}, new dashed edges are introduced to locate vacuum gravitational degrees of freedom. An old matter diagram $\Gamma$ has its edges broken into smaller pieces to become the new subgraph $\gamma_\Gamma$ with all the solid edges. The probability amplitude $A$ for a matter scalar field process on quantum spacetime reads \cite{JiaWorldFunction}
\begin{align}\label{eq:mfp}
&A = \lim_{\text{sk.}} \sum_{g} \sum_\Gamma \frac{V[\gamma_\Gamma]}{N[\gamma_\Gamma,\sigma]}
\prod_{h\notin \gamma_\Gamma} H_h
\prod_{k\in \gamma_\Gamma} D_k,
\\
&H_h=\frac{\sigma_h}{\sigma_h-6i \alpha_h  s_h \Delta_h^{-1} \ln \Delta_h},
\label{eq:pge}
\\
& D_k=\Delta_k^{1-3\xi} \int_0^\infty \frac{dl_k}{(4\pi i l_k)^{2}}
\exp\{i[\sigma_k/2
\nonumber
\\
& \qquad \qquad \qquad \qquad - 3 i \alpha_k s_k \Delta^{-1}_k \ln \Delta_k]/l_k-im^2l_k\},
\label{eq:mge}
\\
&s_j = \abs{2\sigma_j}^{1/2},
\label{eq:gd}
\\
&\alpha_j=\overline{\alpha} \sum_{\{l,m,n\}}\frac{s_ls_ms_n}{N},
\label{eq:alpha}
\\
&\Delta_j(s_j,\rho_j) = \sinc^{-3} \bigg(s_j\sqrt{\frac{\rho_j}{3}}\bigg),
\label{eq:vvmd}
\\
&\sum_{g}=
\prod_j \int_{-\infty}^{\infty} d\rho_j \int_{-a(\rho_j)^2/2}^{a(\rho_j)^2/2} d\sigma_j, 
\quad a(\rho)=
\begin{cases}
\pi \sqrt{\frac{3}{\rho}}, \quad \rho>0,
\\
\infty, \quad \rho\le 0.
\end{cases}
\label{eq:pis}
\end{align}
Here a physical configuration is a skeleton graph (e.g., right of \Cref{fig:wqgpic}) with two numbers $\sigma, \rho\in \mathbb{R}$ (physical meaning explained below) assigned to each edge. $\sigma$ is quadratic in length dimension, while $s$ in (\ref{eq:gd}) is the corresponding non-negative geodesic distance linear in length dimension. Dashed edges labelled by $h$ are assigned \textit{pure gravity} amplitudes $H_h$, and solid edges labelled by $k$ are assigned \textit{matter-gravity} amplitudes $D_k$. The world functions $\sigma_k$ on solid edges implicitly contain an infinitesimal imaginary part $i\epsilon$ (as in the Feynman prescription) to make the $l_k$-integral in (\ref{eq:mge}) converge. In (\ref{eq:alpha}), $\overline{\alpha}$ is the gravitational coupling constant (inversely proportional to Newton's constant), and the average is over all $N$ many sets of 3 edges sharing a same vertex with $j$. $\xi$ is the matter-gravity coupling constant as appears in $(\square+m^2+\xi R)\phi(x)=0$. 
$\sum_\Gamma$ sums over matter Feynman diagrams $\Gamma$ as topological classes of matter configurations. For example, \Cref{fig:wqgpic} shows the one-loop $\Gamma$ with $\gamma_\Gamma$ as its corresponding subgraph on the WQG skeleton on the right. $\sum_\Gamma$ means for each different $\Gamma$ such as the tree, one-loop or two-loop graph, we pick \textit{one} corresponding $\gamma_\Gamma$ subgraph on the WQG skeleton to include in the sum. $V[\gamma_\Gamma]$ is the same vertex factor as in the matter Feynman rules for $\Gamma$. $N[\gamma_\Gamma,\sigma]$ is a generalization of the Feynman diagram symmetry factor to count the number of skeleton graph relabelling that preserves the matter subgraph $\gamma$, the $(\sigma,\rho)$-configuration on the edges, and $V[\gamma_\Gamma]$ on the vertices. The path integral sum over physical configurations $\sum_{g}$ is specified by (\ref{eq:pis}). $\lim_{\text{sk.}}$ indicates that a particular skeleton graph gives an approximation to the physical amplitude $A$, the exact value of which is approached by going to ever larger skeleton graphs.

To connect with familiar physics, we take away terms proportional to the gravitational coupling $\overline{\alpha}$ for a moment. Pure gravity edges become $H_h=1$ and are now trivial. Matter-gravity amplitudes become $D_k=\Delta_k^{1-3\xi} \int_0^\infty \frac{dl_k}{(4\pi i l_k)^{2}}
\exp{i\sigma_k/2l_k-im^2l_k}$, which coincides with the Schwinger proper time representation \cite{Schwinger1951OnPolarization} of the Feynman propagator for a massive scalar field coupled to gravity with coupling constant $\xi$ in the ``one-segment'' path integral approximation \cite{Bekenstein1981Path-integralSpacetimeb}. This expression would lead to the usual UV divergences of ordinary QFT. Formula (\ref{eq:mfp}) introduces quantum gravitational effects to matter QFT through 1) the path integral sum $\sum_{g}$ over spacetime configurations, and 2) the $\alpha$-dependent terms in $H$ and $D$. \textit{We will see below that these two aspects together eliminate the UV divergences.}

Regarding the spacetime configurations, the world function $\sigma$ encodes the causal structures and spacetime distances. An additional variable $\rho$ is needed to incorporate information on the second derivative of $\sigma$, analogous to needing momentum in addition to position in mechanics. Physically
\begin{align}\label{eq:rho}
\rho=\bar{\sigma}^2-\omega^2+R_{ab}u^a u^b
\end{align}
encodes information on curvature $R_{ab}$ (Ricci tensor), shear $\bar{\sigma}$ and rotation $\omega$. Formula (\ref{eq:rho}) has its root \cite{JiaWorldFunction} in the Raychaudhuri equation \cite{Poisson2004AToolkit, Abreu2011SomeEquation}, where $u^a$ is the unit tangent vector along the geodesics of the skeleton edges. $\rho$ is assumed constant on each edge, just like in evaluating the path integral for a point particle, the velocity is assumed constant on the segments of the piecewise linear trajectories. For $\rho>0$, the $\sigma$-integral terminates at $a(\rho)$ as indicated in \Cref{eq:pis} to avoid geodesics running into each other to form caustics \cite{JiaWorldFunction}. We call this the \textbf{caustic constraint}.

The $\alpha$-dependent terms in $H$ and $D$ originate from an analogue of the Einstein-Hilbert action on the skeletons. While a detailed derivation can be found in \cite{JiaWorldFunction}, the basic idea is to 1) replace $\overline{\alpha} dx^4 \sqrt{-g} R$ by $\overline{\alpha} s_j s_l s_m s_n \Delta^{-1} R$ using the correspondence between $\sqrt{-g}$ and the Van Vleck-Morett determinant $\Delta$ \cite{VanVleck1928TheMechanics., Morette1951OnIntegrals, Visser1993VanSpacetimes}, and then 2) apply Parker's path integral correspondence formula \cite{Parker1979PathSpace} \begin{align}\label{eq:pmf2}
\exp{i(\frac{\sigma}{2l}+c Rl)} \xleftrightarrow{\sum_\text{path}} \Delta^{3c} \exp{i\frac{\sigma}{2l}}.
\end{align} to eliminate $R(g_{ab})$ in favour of $\Delta(s,\rho)$. The form of $\Delta$ in (\ref{eq:vvmd}) is derived under the assumption that $\rho$ is constant on each edge \cite{JiaWorldFunction}.

This finishes our short presentation on WQG. As a final word, we stress that the use of discrete skeleton graphs in general \textit{neither implies that spacetime is fundamentally discrete} (e.g., the path integral for a point particle on continuum spacetime is standardly evaluated using discretized paths), \textit{nor imposes a regularization} (e.g., Feynman loop diagrams diverge without extra regularization even though the diagrams are discrete graphs). We shall see below that path integrating the continuous spacetime distances over different causal structures provides a regularization.


\section{Hardy sum}

The theory is UV finite because the gravitational path integral (\ref{eq:pis}) sums across different causal structures with different signs of $\sigma$. For an arbitrary function $f(\sigma)$,
\begin{align}
&\int_{-a^2/2}^{a^2/2} d\sigma f(\sigma)
\nonumber
\\
=& 
\int_{-a^2/2}^0 d\sigma 
f(\sigma=-s^2/2)
+
\int_{0}^{a^2/2} d\sigma f(\sigma=s^2/2)
\label{eq:hs1}
\\
=&
\int_{a}^0 (-s ds) ~ f(\sigma=-s^2/2)
+
\int_{0}^a (s ds) ~ f(\sigma=s^2/2)
\label{eq:hs2}
\\
=&\int_{0}^a ds ~ 
s[
f(\sigma=-s^2/2)
+
f(\sigma=s^2/2)
].
\label{eq:hsi}
\end{align}
We call $f_{\%}(s):=f(\sigma=-s^2/2)+f(\sigma=s^2/2)$ the \textbf{Hardy sum} of $f(\sigma)$, and $s f_{\%}(s)$ the \textbf{Hardy sum integrand} \footnote{The symbol ``$\%$'' is a pictorial representation of combining timelike and spacelike contributions (two circles) separated by the lightcone (tilted line). The expressions are named in honor of Lucien Hardy, who pioneered the study of indefinite causal structure as a central feature for quantum gravity \cite{HardyProbabilityGravity, *hardy2007towards}.}. There are two channels for potential divergence cancellation, from the sum over causal structures in the Hardy sum, and from the extra factor of $s$ in the integrand.  

The identity (\ref{eq:hsi}) can be applied to the path integral sum over $\sigma$ in (\ref{eq:pis}). Let $A=s^2/2$, $B=6 \alpha  s \Delta^{-1} \ln \Delta$.
For $H$,
\begin{align}
H_{\%}=&
\frac{-s^2/2}{-s^2/2-6i \alpha  s \Delta^{-1} \ln \Delta}
+
\frac{s^2/2}{s^2/2-6i \alpha  s \Delta^{-1} \ln \Delta}
\\
=&
\frac{-A}{-A-i B}
+
\frac{A}{A-i B}
\\
=&
\frac{2A^2}{A^2+B^2}.
\label{eq:hge}
\end{align}
For $D$, what we care about is the UV regime where mass is negligible. Denote the massless version of $D$ by $\overline{D}$. By $\int_0^\infty \frac{dl}{l^2}\exp{i\frac{X+i\epsilon}{l}} = i/(X+i\epsilon)$ (easily verified by changing variable from $l$ to $1/l$),
\begin{align}
\overline{D}=&\Delta^{1-3\xi} \int_0^\infty \frac{dl}{(4\pi i l)^{2}}
\exp{i[\sigma/2- 3 i \alpha s \Delta^{-1} \ln \Delta]/l}
\label{eq:mme}
\\
=&\frac{1}{-8\pi^2} \frac{i \Delta^{1-3\xi}}{\sigma-6i \alpha  s \Delta^{-1} \ln \Delta}.
\label{eq:mem}
\\
\overline{D}_{\%}
=&
\frac{1}{-8\pi^2} (
\frac{i \Delta^{1-3\xi}}{-s^2/2-6i \alpha  s \Delta^{-1} \ln \Delta}
\nonumber
\\
& \qquad \qquad \qquad +\frac{i \Delta^{1-3\xi}}{s^2/2-6i \alpha  s \Delta^{-1} \ln \Delta}
)
\\
=&
\frac{1}{-8\pi^2} (
\frac{i \Delta^{1-3\xi}}{-A-iB}
+
\frac{i \Delta^{1-3\xi}}{A-iB}
)
\\
=&
\frac{1}{4\pi^2} \frac{B \Delta^{1-3\xi}}{A^2+B^2}.
\label{eq:hme}
\end{align}
We will use these expressions to analyze the UV structure of the theory. 

Before this, we note incidentally that the Hardy sum enables efficient computations of the WQG amplitude (\ref{eq:mfp}) for vacuum gravity and for gravity with massless matter. This is because the Monte Carlo method applies without the ``sign problem'' when $H_{\%}\ge 0$, and when $\overline{D}_{\%}\ge, \le 0$ for $\rho\ge, <0$, respectively \footnote{By (\ref{eq:vvmd}), $\Delta=\sinc^{-3}(s\sqrt{\frac{\rho}{3}})$. For $\rho> 0$, when the caustic constraint is obeyed $s\sqrt{\frac{\rho}{3}}\in [0,\pi)$, so $\Delta\ge 1$. Then $\ln \Delta\ge 0$, and $\overline{D}_{\%}\ge 0$. For $\rho \le 0$, $s\sqrt{\frac{\rho}{3}}$ is purely imaginary with a non-negative imaginary part. For $x\ge 0$, $\sinc(ix)=\sinh(x)/x\ge 1$, so $0<\Delta\le 1$. Then $\ln \Delta\le 0$, and $\overline{D}_{\%}\le 0$. In both cases $H_{\%}\ge 0$ by (\ref{eq:hge}).}. This remarkable opportunity to compute Lorentzian quantum gravitational path integrals with many degrees of freedom efficiently without performing a ``Wick rotation'' will be explored elsewhere. 


\section{Ultraviolet finiteness}


To investigate the UV regime, we study the $s\rightarrow 0$ limit. We will see that in contrast to ordinary QFT, this short distance limit is finite when quantum gravitational effects are taken into account. 

On a pure gravity edge, we have
\begin{align}
\lim_{s\rightarrow 0} H=1,\\
\lim_{s\rightarrow 0} H_{\%}=2,\\
    \lim_{s\rightarrow 0} s H_{\%}=0.\label{eq:geil}
\end{align}
These are derived by applying L'Hopital's rule to (\ref{eq:pge}), (\ref{eq:hge}), and $s H_{\%}$ for the case $\rho<\infty$ under the assumption that $s$ on adjacent edges are finite (reasonable in the UV regime) so that $\alpha<\infty$. For $\rho\rightarrow \infty$, due to the caustic constraint $s\sqrt{\frac{\rho}{3}}<\pi$ we need to fix $x=s\sqrt{\frac{\rho}{3}}$ for some $0<x<\pi$ in taking the $\rho\rightarrow \infty$ and $s\rightarrow 0$ limits. Then $\Delta= \sinc^{-3}(x)>1$ as well as $B$ are finite constants, and
\begin{align}\label{eq:pgl}
\lim_{s\rightarrow 0} H=
\lim_{s\rightarrow 0} H_{\%}=
\lim_{s\rightarrow 0} s H_{\%}=0.
\end{align}
Unlike graviton propagators, pure gravity edges create no UV divergences in this non-perturbative approach.

On a matter-gravity edge,
\begin{align}
\lim_{s\rightarrow 0} s \overline{D}_{\%}
=
\lim_{s\rightarrow 0} \frac{3 \alpha}{2\pi^2} \frac{s^2}{s^2}  \frac{\Delta^{-3\xi} \ln \Delta}{\frac{s^2}{4}+(6\alpha \Delta^{-1}\ln \Delta)^2},
\label{eq:mil}
\end{align}
\begin{align}
&\lim_{s\rightarrow 0} \frac{\Delta^{-3\xi} \ln \Delta}{\frac{s^2}{4}+(6\alpha \Delta^{-1}\ln \Delta)^2}
=
\lim_{s\rightarrow 0} \frac{(\Delta^{-3\xi} \ln \Delta)'}{\frac{s}{2}+[(6\alpha \Delta^{-1}\ln \Delta)^2]'}
\label{eq:ld1}
\\
&=
\lim_{s\rightarrow 0} \frac{(\Delta^{-3\xi} \ln \Delta)''}{\frac{1}{2}+[(6\alpha \Delta^{-1}\ln \Delta)^2]''}
=\frac{2\rho}{3},
\label{eq:ld3}
\\
&\implies \lim_{s\rightarrow 0} s \overline{D}_{\%}
=\frac{\rho \alpha}{\pi^2}.
\label{eq:ld4}
\end{align}
The finite value in (\ref{eq:ld4}) is derived assuming $\alpha, \rho<\infty$. Here prime denotes derivative with respect to $s$. One can check that the $s\rightarrow 0$ limits of the numerators and denominators on both sides of (\ref{eq:ld1}) are $0$, so we applied L'Hopital's rule twice to obtain the result $2\rho/3$. When $\alpha<0, \rho\rightarrow \infty$, we again we fix $x=s\sqrt{\frac{\rho}{3}}$ for some $0<x<\pi$ due to the caustic constraint. Then $\Delta= \sinc^{-3}(x)>1$ is a finite constant, and (\ref{eq:mil}) yields $\lim_{s\rightarrow 0} s D_{\%}=[24 \pi ^2 \alpha  \sinc^{6-9\xi}(x) \ln \left(\sinc^{-3}(x)\right)]^{-1}$, which is finite.
Therefore we have UV finiteness for all cases.

Some remarks regarding the matter-gravity edges: 1) 
Quantum gravitational effects come in at two places -- the additional $\alpha$-dependent term coming from the Einstein-Hilbert action, and the summation $\sum_{g}$ coming from the gravitational path integral. \textit{Both are needed for the UV finiteness.} Without the extra $s$ factor from the $\sigma$, $\lim_{s\rightarrow 0} \overline{D} = \pm i \infty$ for $\sigma=\mp s^2/2$, as seen from (\ref{eq:mem}). Without the $\alpha$-dependent term $\overline{D}$ multiplied by $s$ still has a pole, as seen from (\ref{eq:mil}).
2) \textit{Performing the hardy sum over causal structures is also necessary for the UV finiteness.} As $s$ approaches $0$, $s\overline{D}$ without the Hardy sum diverges, as seen from (\ref{eq:mem}). 
3) The $\alpha$-dependent modification term $-iB/2$ satisfies $\lim_{s\rightarrow 0} B/s=0$, $\lim_{s\rightarrow 0} (B/s)'=0$, and $\abs{\lim_{s\rightarrow 0} (B/s)''}<\infty$. Some other form of modification obeying the same conditions will also allow the steps (\ref{eq:ld1}) to (\ref{eq:ld3}) carry through to reach a finite limit.

\section{Singularity resolution}

A quantum gravitational path integral sums over spacetime configurations, including singular ones. What does it mean to resolve singularity in this setting? We will show that singular configurations have zero amplitudes so do not contribute to the WQG path integral. In this sense singularities are resolved. 


When $s\rightarrow 0$ on a gravity edge, its Hardy sum integrand approaches zero by (\ref{eq:geil}) and (\ref{eq:pgl}). This implies that the amplitude for the whole spacetime configuration approaches zero as long as there is at least one gravity edge on which $s\rightarrow 0$. Singular spacetimes fulfil this condition. With $\rho=\bar{\sigma}^2-\omega^2+R_{ab}u^a u^b$ as in (\ref{eq:rho}), divergent curvature corresponds to $\rho\rightarrow \infty$ on some edge(s). The caustic constraint (\ref{eq:pis}) implies $s\rightarrow 0$.
Alternatively, the geodesic incompleteness criterion for singularities (used e.g, in Penrose-Hawking singularity theorems \cite{Hawking1973TheSpace-Time, wald2010general}) says geodesics cannot extend beyond a certain distance in some regions on a singular spacetime configuration. In the present setting this translates to bounds on $s$ on the edges. Due to $\lim_{\text{sk.}}$ in (\ref{eq:mfp}), the physical amplitude is approached by increasing the number of edges indefinitely, so $s\rightarrow 0$ on some edge(s). In particular, there will be some gravity edge(s) on which $s\rightarrow 0$ as we introduce gravity edges in the limiting procedure.

To obtain a vanishing amplitude, it is not necessary to perform the Hardy sum since both $\lim_{s\rightarrow 0} s H_{\%}$ and $\lim_{s\rightarrow 0} sH$ (for $\sigma\ge 0$ or $\sigma\le 0$) are zero. Nevertheless we regard the singularity resolution as a consequence of indefinite causal structure, because the crucial factor $s$ comes from (see the step from (\ref{eq:hs1}) to (\ref{eq:hs2})) the $\sigma$-integral, which achieves the sum over causal structures.

Incidentally, generating matter worldline loops (self-crossing worldlines) also requires some gravity edge(s) to have $s=0$. Hence not only singularity configurations, but also configurations with such matter loops are suppressed.


\section{Generality}

We have demonstrated using the World Quantum Gravity approach that summing over causal structures eliminates UV divergences and resolves spacetime singularities. How general is this mechanism of UV regularization by indefinite causal structure?

As mentioned in the introduction, indefinite causal structure is generically present in path integral formulations of quantum gravity. What we need is a model-independent way to study its effect on UV physics. For this we turn to information-theoretic formulations of quantum theory and consider model-independent statistical correlations. In the above analysis of UV finiteness and singularity resolution, the crucial input is that the amplitudes drop to zero as the invariant distance $s$ approaches zero. Using the framework of process matrices \cite{oreshkov2012quantum} that incorporates quantum causal structures, it was found in \cite{jia2017quantum, jia2018reduction} that indefinite causal structure reduces correlations generically. In particular, for the largest causal structure fluctuations with comparable weights assigned to different causal structures, the quantum correlations measured by coherent information are reduced to zero. 

In quantum gravity the regime of large causal structure fluctuations is the regime of large quantum gravitational fluctuations at short distance scales. If we infer that the amplitudes vanish as the quantum correlations vanish, then the model-independent result hints that indefinite causal structure leads to UV regularity generically. It will be very interesting to check explicitly if the suggestion is realized in different approaches to quantum gravity.

\section*{Acknowledgement}

I am very grateful to Lucien Hardy and Achim Kempf for discussion, guidance and support.

Research at Perimeter Institute is supported in part by the Government of Canada through the Department of Innovation, Science and Economic Development Canada and by the Province of Ontario through the Ministry of Economic Development, Job Creation and Trade. This work is made possible through the support of the grant ``Operationalism, Agency, and Quantum Gravity'' from FQXi. The opinions expressed in this work are those of the author and do not necessarily reflect the views of the funding agencies. 

\bibliography{mendeley.bib}

\end{document}